\documentclass{basi}
\usepackage{graphicx}
\usepackage{mathptmx}
\newcommand{\Hi}{\textsc{Hi}}
\newcommand{\subHi}{H{\scriptscriptstyle I}}
\newcommand{\la}{\,\rlap{\raise 0.5ex\hbox{$<$}}{\lower 1.0ex\hbox{$\sim$}}\,}
\pubyear{2006}
\volume{00}
\pagerange{\pageref{firstpage}--\pageref{lastpage}}
\begin{document}

\title[GMRT limits on cool H{\scriptsize\it I} in galaxy groups]{Giant
Metrewave Radio Telescope limits on cool H{\large\bf I} in galaxy groups}

\author[D.A.\ Green et al.]{D.A.\ Green,$^1$ M.S.\ Clemens,$^2$
        P.\ Alexander,$^1$ H.\ Cullen$^1$ and B.\ Nikolic$^1$\\
  $^1$Mullard Radio Astronomy Observatory, Cavendish Laboratory,
      19 J.~J.~Thomson Avenue,\\\ Cambridge CB3 0HE, United Kingdom\\
  $^2$Dipartimento di Astronomia, Universita degli Studi di Padova,
      Vicolo dell'Osservatorio, 2,\\\ 35122 Padova, Italy}

\date{Received 2006 November 14th}

\maketitle

\label{firstpage}

\begin{abstract}
We present Giant Metrewave Radio Telescope 21-cm {\Hi} observations towards a
sample of compact radio sources behind galaxy groups, to search for {\sl cool}
{\Hi}. The results -- from high dynamic range spectra for 8 lines-of-sight
through 7 galaxy groups -- do not show any evidence for absorption by cool
{\Hi}. At a resolution of 20 km s$^{-1}$, the optical depth upper limits
obtained were between 0.0075 and 0.035 ($3\sigma$); these correspond to upper
limits of a few times $10^{23}$ m$^{-2}$ for the column density of any cool
{\Hi} along these lines of sight (assuming a spin temperature of 100 K).
\end{abstract}


\section{Introduction}

A central question in understanding galaxy evolution is how primordial gas
clouds first evolve into star-forming systems. There are good reasons to expect
that gas which has never been associated with star-forming galaxies still
exists in the Universe today. In cold dark matter cosmologies, low-mass
structures (dark matter halos and baryonic matter) are statistically the first
to form. If the baryonic matter cools it will form bound clouds and structure
formation proceeds hierarchically. The mass spectrum expected for primordial
clouds is steep (with the number varying with mass, $m$, as $n(m) \propto
m^{-2}$), and if the smaller clouds do not all merge, or are disrupted, such
clouds may be expected to survive to the present epoch. The most likely regions
in which to find such primordial extragalactic clouds are in groups dominated
by late-type galaxies, for several reasons. First, the collision rate in a
galaxy group is expected to be lower than in a cluster, and therefore we might
expect that low-mass clouds in galaxy groups have survived to the present day.
Second, it is observed that groups rich in elliptical galaxies often show
extended X-ray emission (as in centrally-condensed clusters), whereas groups
dominated by late-type galaxies never show extended X-ray emission (Mulchaey et
al.\ 1996). Nath \& Chiba (1995) have suggested that this X-ray emitting gas
has its origin in the disruption of low-mass, primordial clouds. This
observation suggests that groups rich in late-type galaxies also contain a
large mass of gas not associated with galaxies. If we interpret this lack of
X-ray emission in groups dominated by late-type galaxies as evidence for the
non-disruption of low-mass clouds, then provided the clouds exceed their virial
temperature (so that they have not collapsed to form stars), and are neutral,
these clouds may be detectable.

Detections of intra-group atomic hydrogen which may be genuinely primordial are
extremely limited, but some groups do show gaseous structures that are hard to
explain as interaction products or outflows. These include, the M96 group
(Schneider, Saltpeter \& Terzian 1986), around NGC 5291 (Malphrus et al.\
1997), the environment of NGC 4532 (Hoffman et al.\ 1999), the Virgo cluster
(Davies et al.\ 2004, see also Walter, Skillman \& Brinks 2005), and near local
group Dwarf galaxies (Bouchard, Carignan \& Staveley-Smith 2006). The recent
HIPASS survey has also revealed a small number of {\Hi} clouds with no optical
counterparts although not specifically in groups (e.g.\ Kilborn et al.\ 2000,
2002; and also see Doyle et al.\ 2005). Several searches for low mass {\Hi}
clouds in galaxy groups have resulted in non-detections down to thresholds as
low as $3 \times 10^{6}$ $\rm M_{\odot}$ (e.g.\ Zwaan \& Briggs 2000; Zwaan
2001; Dahlem, Ehle \& Ryder, 2001; de Blok et al.\ 2002).

Here we present Giant Metrewave Radio Telescope (GMRT; see Rao 2002) 21-cm
observations to test whether (primordial) intra-group {\Hi} is common in groups
which might in some sense be considered young. If the gas is cool and/or in
low-mass clouds -- as might be expected for a steep $n(m)$ spectrum -- then it
is likely that a lower mass detection limit could be reached by searching for
absorption rather than emission from the neutral gas. Other searches for
primordial {\Hi} have been, and indeed are being carried out, usually at high
redshifts (e.g.\ Bebbington 1986; Wieringa, de Bruyn \& Katgert 1992;
Taramopoulos, Briggs \& Turnshek 1994; Weintroub et al.\ 1999), but these have
only probed for very massive primordial structures. A similar experiment to
ours has been carried out by Kanekar \& Chengalur (2003) with the GMRT, but
this was aimed at detecting {\Hi} absorption towards damped Lyman alpha systems
(see also Kanekar \& Chengalur 2005). Details of our observations towards
compact continuum sources behind groups of galaxies, which place strong limits
on the existence of low-mass clouds in the local Universe, are given in
Section~\ref{s:data}, with the results and conclusions given in
Sections~\ref{s:results} and \ref{s:conclusions}

\section{GMRT Observations and Data Reduction}\label{s:data}

\subsection{Selected Fields}

To maximise our sensitivity to {\sl cool} atomic gas, we search for {\Hi} in
{\em absorption} towards background radio sources. Our sample consists of seven
groups (from the catalogues of Garcia 1993 and Ramella, Pisani \& Geller 1997),
as shown in Table~\ref{t:sample}. These satisfy the following criteria. (i) The
group is dominated by late-type galaxies. It is likely that such groups
represent environments where fewer galaxy--galaxy interactions have taken
place, and hence primordial gas is more likely to survive, and also such groups
do not have X-ray emission which may originate from disrupted, low-mass
primordial clouds. (ii) The group is away from the Galactic plane. (iii) There
is one ore more compact continuum source brighter than 60~mJy at 1.4~GHz
visible in the NVSS survey (Condon et al.\ 1998) close to the centre of the
group, towards which 21-cm absorption measurements can be made. Given the
selected galaxy groups are nearby, with velocities less than $\sim 10^4$ km
s$^{-1}$, then it is reasonable to expect all the selected, compact NVSS
sources to be more distant than the groups.

\subsection{Observations and Data Reduction}

These fields were observed in two sessions with the GMRT, 2003 January 6th and
7th, and October 17th. Each group was observed for between 1 and 2 hours each
(depending on scheduling constraints), with appropriate correlator frequency
settings. The observations were made using 128 spectral channels, with a total
bandwidth of usually 4~MHz, or in some cases 8~MHz (e.g.\ to allow two targets
to be observed with the same reference velocity, to optimise calibration time,
or for a group with a high velocity dispersion), giving channels of width 6.6
or 13.2 km s$^{-1}$ respectively. The data were observed at a fixed frequency,
without Doppler correction for diurnal variations, which is small for the
duration of the observations compared with the individual channel width.
Independent right and left Stokes parameter spectral data were recorded.
Bright, primary, calibrators 3C48, 3C147 or 3C286 were observed for each of the
velocity setting used. Secondary calibrator sources close to each of the galaxy
group were observed for $\approx 4$ min every $\approx 30$ min. During these
observations typically 26 of the 30 antennas of the GMRT were available.

\begin{table*}
\centering
\begin{minipage}{12.5cm}
\caption{Observed galaxy groups. R $=$ from Ramella et al.\ (1997); LGG $=$
from Garcia (1993).}\label{t:sample}
\smallskip
\tabcolsep4pt
\begin{tabular}{ccccccc}\hline
  \multicolumn{3}{c}{Group}                 & \multicolumn{2}{c}{Position}     & Observed  & Date          \\
  ID       &   velocity     &  dispersion   & \multicolumn{2}{c}{J2000}        & Bandwidth & Observed      \\
           &  / km s$^{-1}$ & / km s$^{-1}$ &  h\quad m    &  $^\circ$\quad$'$ &           &               \\ \hline
   R 97    &      6568      &      139      &    10 47.3   &     $+$26 19.0    &   4 MHz   & 2003 Jan 6th  \\
  R 110    &     10470      &      800      &    10 59.2   &     $+$10 00.5    &   8 MHz   & 2003 Jan 7th  \\
  R 151    &      6265      &      261      &    11 43.1   &     $+$10 23.0    &   4 MHz   & 2003 Jan 6th  \\
  R 168    &      4790      &      163      &    11 58.5   &     $+$25 08.4    &   4 MHz   & 2003 Oct 17th \\
  R 282    &      6873      &      668      &    14 06.0   &     $+$09 08.0    &   8 MHz   & 2003 Jan 6th  \\
  R 331    &      6905      &      209      &    15 06.7   &     $+$12 44.3    &   8 MHz   & 2003 Jan 6th  \\
 LGG 413   &      3315      &       --      &    17 52.5   &     $+$24 29.9    &   4 MHz   & 2003 Oct 17th \\ \hline
\end{tabular}
\end{minipage}
\end{table*}

The data were calibrated using standard procedures using classic {\sl AIPS}
(Bridle \& Greisen 1994). A few channels near the centre of the band were
inspected, and obvious interference was flagged. These channels were then
averaged, and the observations of the secondary calibrators were then used to
measure the antenna-based amplitude and phase variations throughout the
observations. The overall flux density scale of the observations was set by the
observations of the primary calibrators, which were also used to determine
antenna-based bandpass corrections. For the January observations, a few
correlator channels were consistently identified as not working, and these are
omitted from the results shown here. The calibrated data for both the
calibrator sources, and the galaxy groups, was inspected, and further
interference was identified and flagged. Images of the galaxy groups were made,
in order to check the positions of continuum sources, and also see if they were
resolved. The GMRT consists of 30 antennas each 45-m in diameter, 12 in a
central region $\approx 1$~km in extent, with the others in three arms,
providing baselines up to $\approx 25$~km. At 1.4~GHz the primary response of
the GMRT is $\approx 0.4$ deg FWHM, and the full resolution of the telescope is
$\approx 2$~arcsec. In some cases it was found that the continuum sources were
resolved at the full resolution of the telescope. {\Hi} spectra towards the
continuum sources were produced using the {\sl AIPS} task {\sc ispec}. This
coherently added the calibrated visibilities together, on a channel-by-channel
basis, using appropriate shifts in position away from the phase centre of the
observations. Limits were placed on the range of baselines in the $uv$-plane
that were averaged together if the source was resolved. Placing $uv$-plane
limits on the data averaged together increases the total flux detected, at the
expense of increased noise, but allows increased signal-to-noise in the spectra
to be obtained. (Ideally the expected noise in a single 6.6 km s$^{-1}$
channel, for a 1 hour integration, is $\approx 1$ mJy, and noise levels close
to this value were obtained, see Table~\ref{t:results}.)

\begin{figure}[tp!]
\centerline{\includegraphics[angle=270,width=13.6cm,clip=]{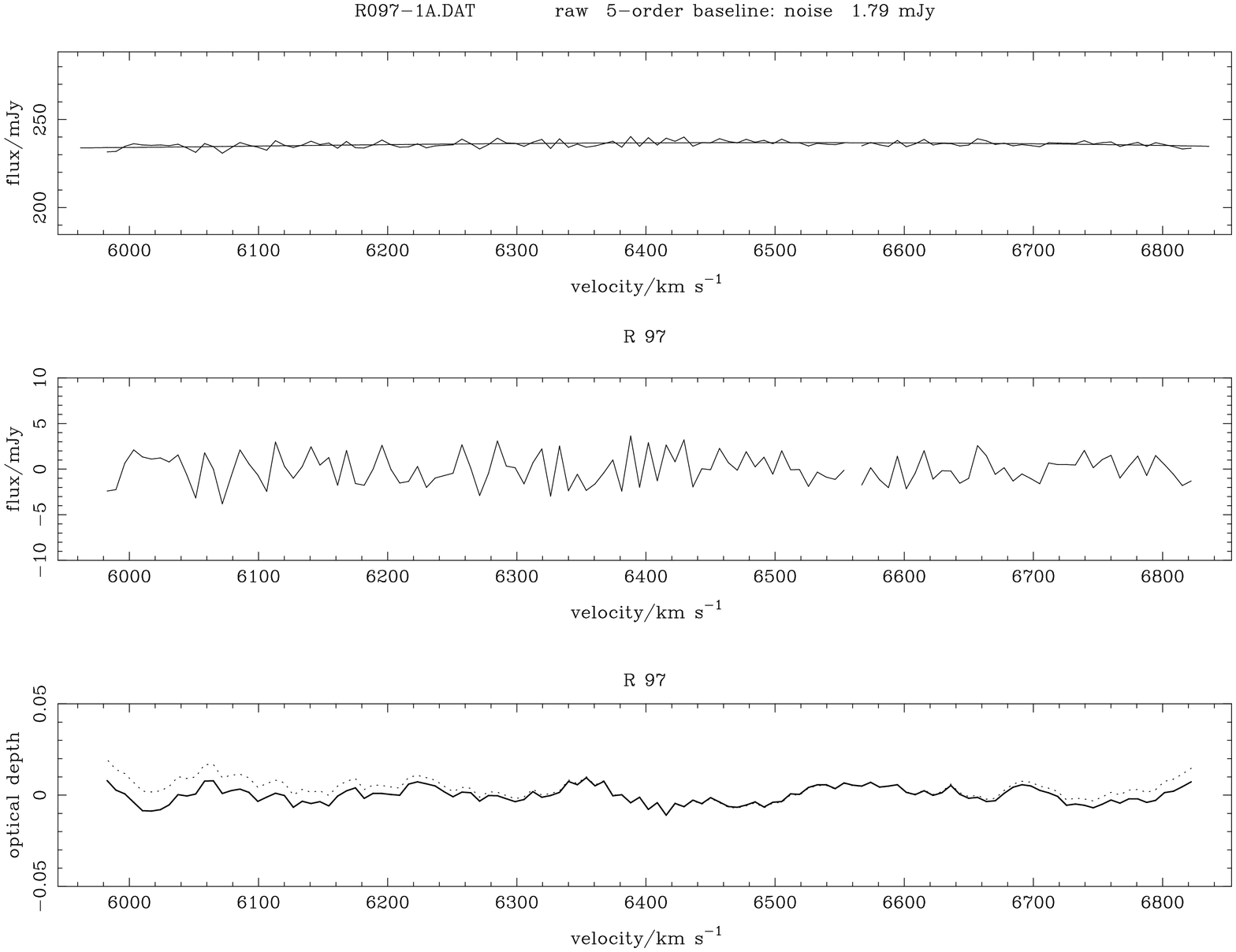}}
  \smallskip
\centerline{\includegraphics[angle=270,width=13.6cm,clip=]{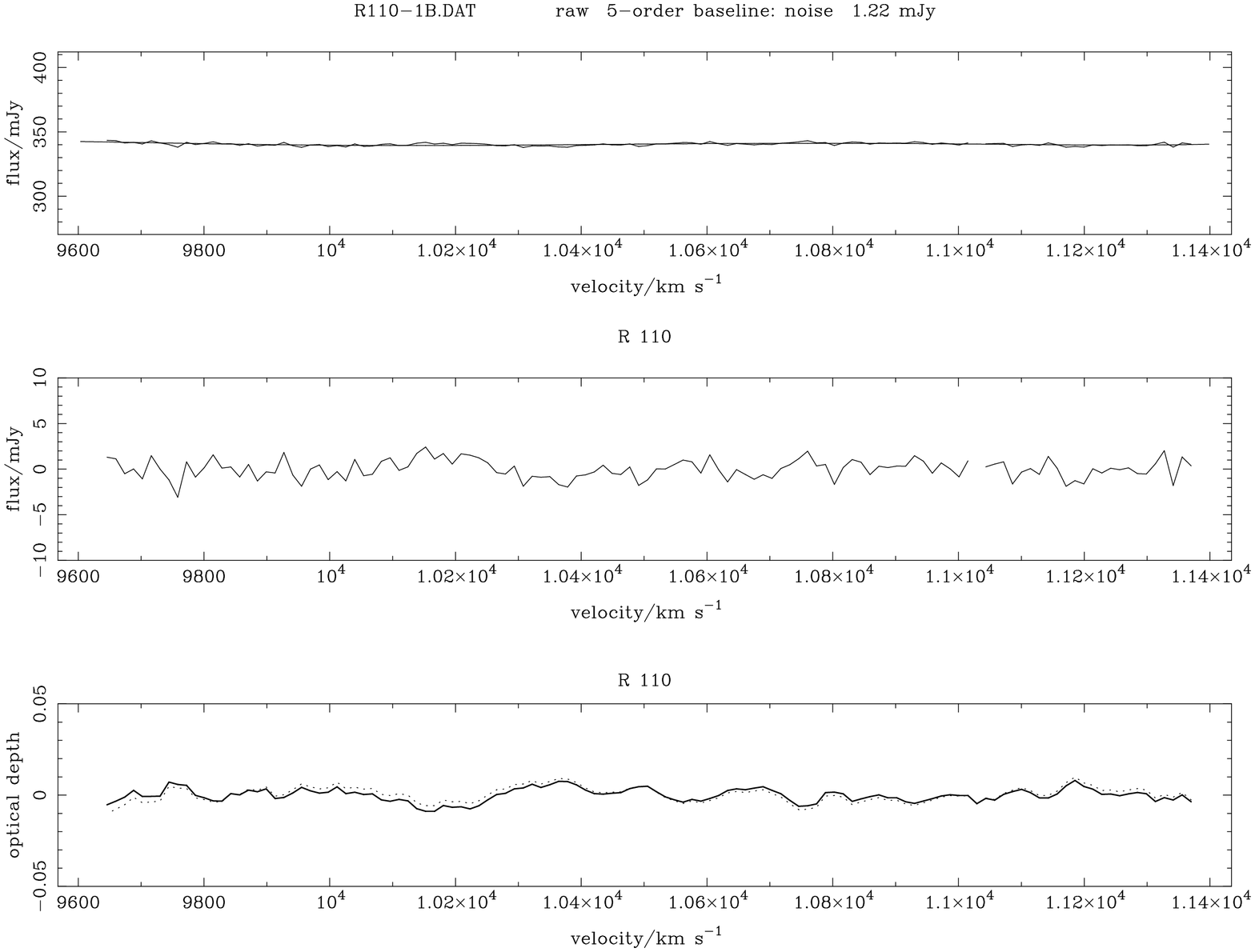}}
  \smallskip
\centerline{\includegraphics[angle=270,width=13.6cm,clip=]{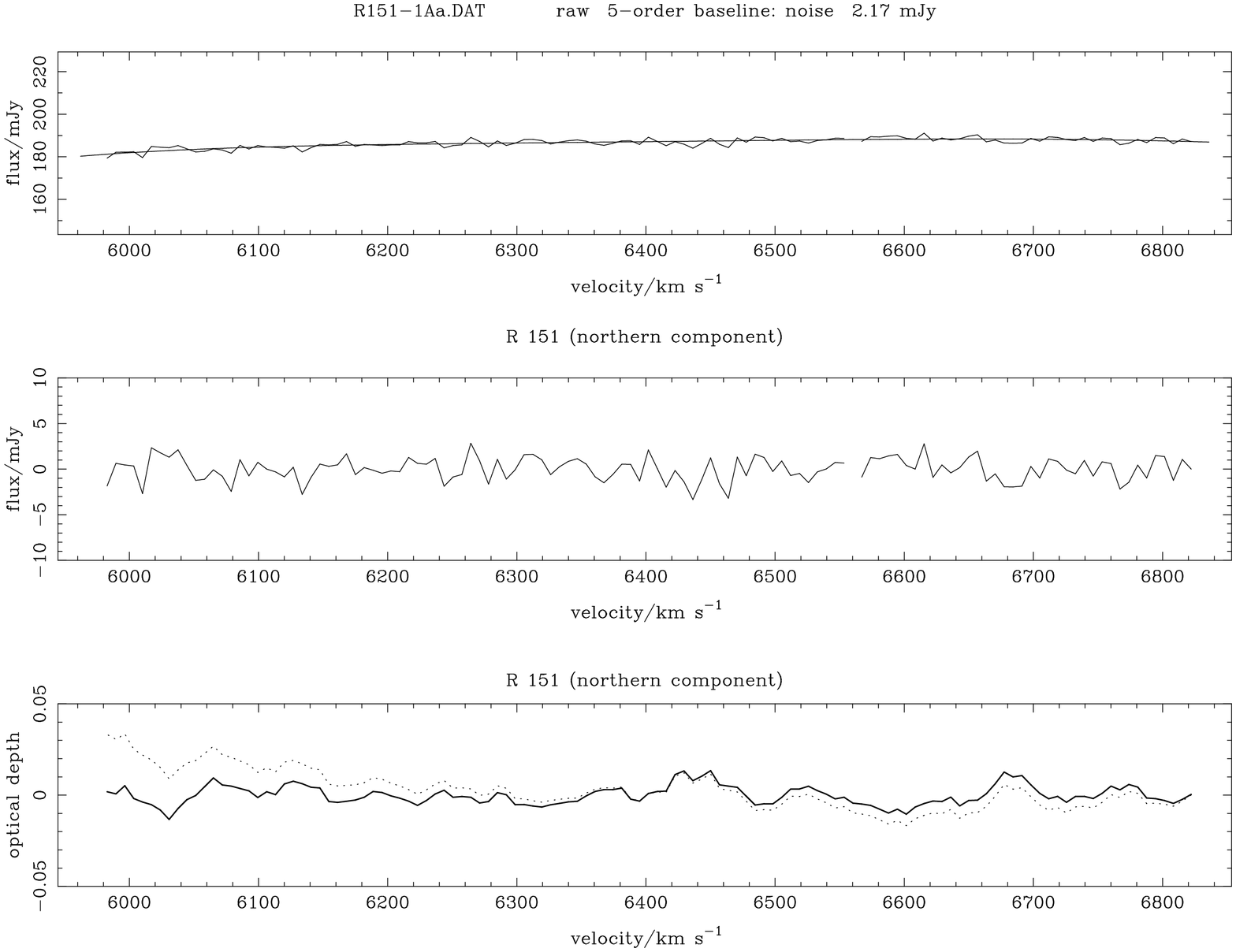}}
  \smallskip
\centerline{\includegraphics[angle=270,width=13.6cm,clip=]{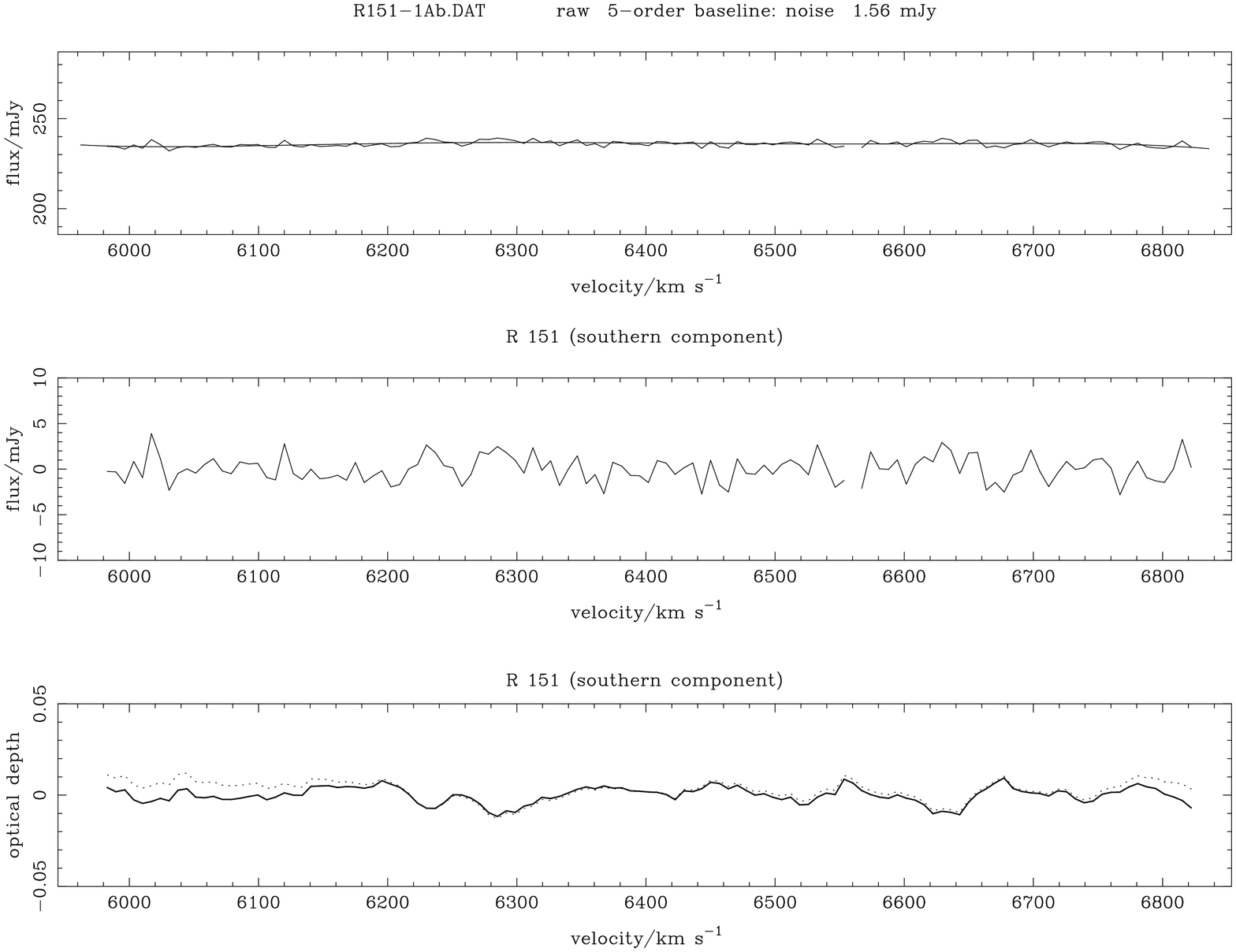}}
  \smallskip
\centerline{\includegraphics[angle=270,width=13.6cm,clip=]{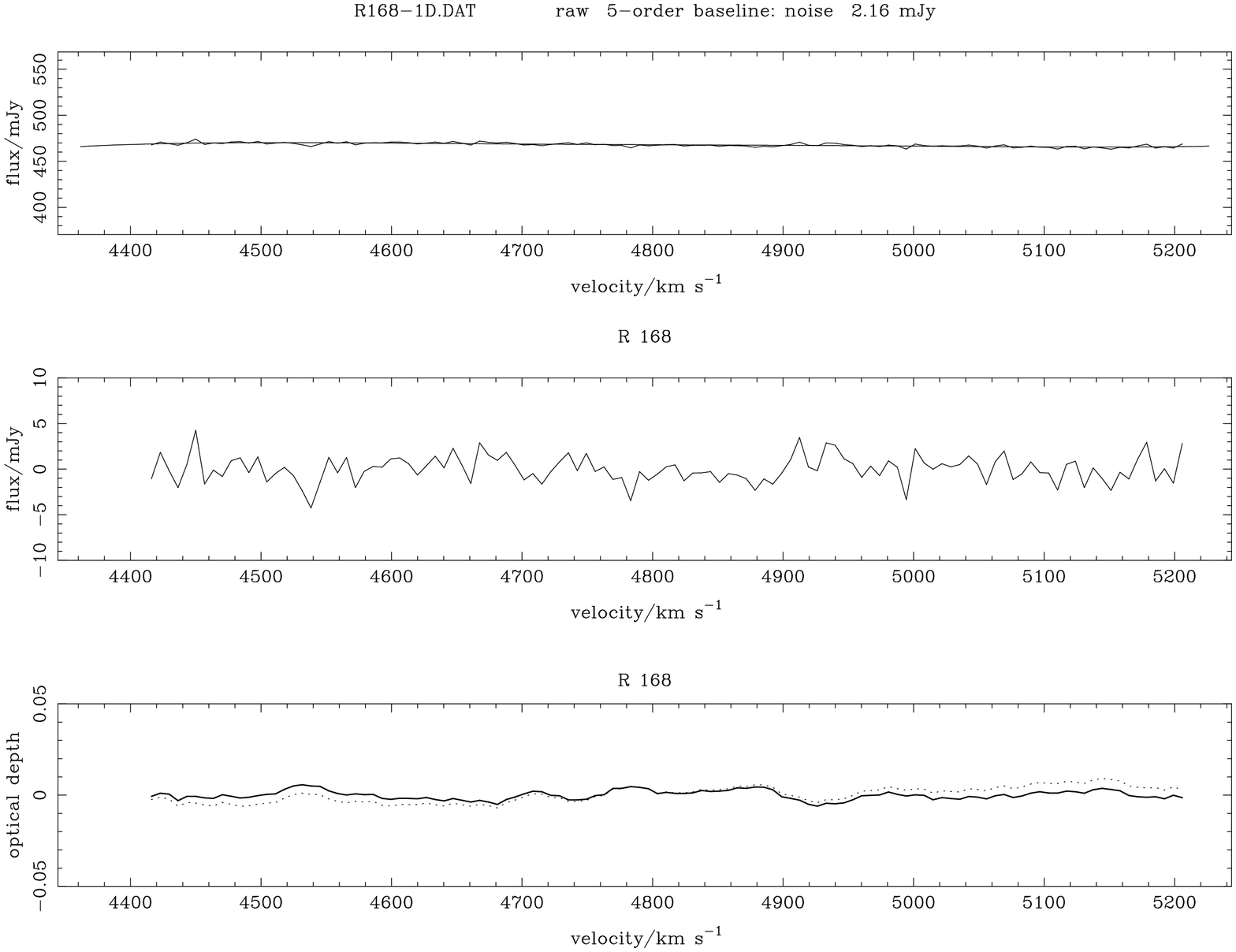}}
\caption{Observed 21-cm {\Hi} absorption spectra, in terms of optical depth,
smoothed to a resolution of 20 km s$^{-1}$, both before (dotted line) and
after (solid line) the removal of a fifth order polynomial
fit.\label{f:results}}
\end{figure}

\addtocounter{figure}{-1}

\begin{figure}[tp!]
\centerline{\includegraphics[angle=270,width=13.6cm,clip=]{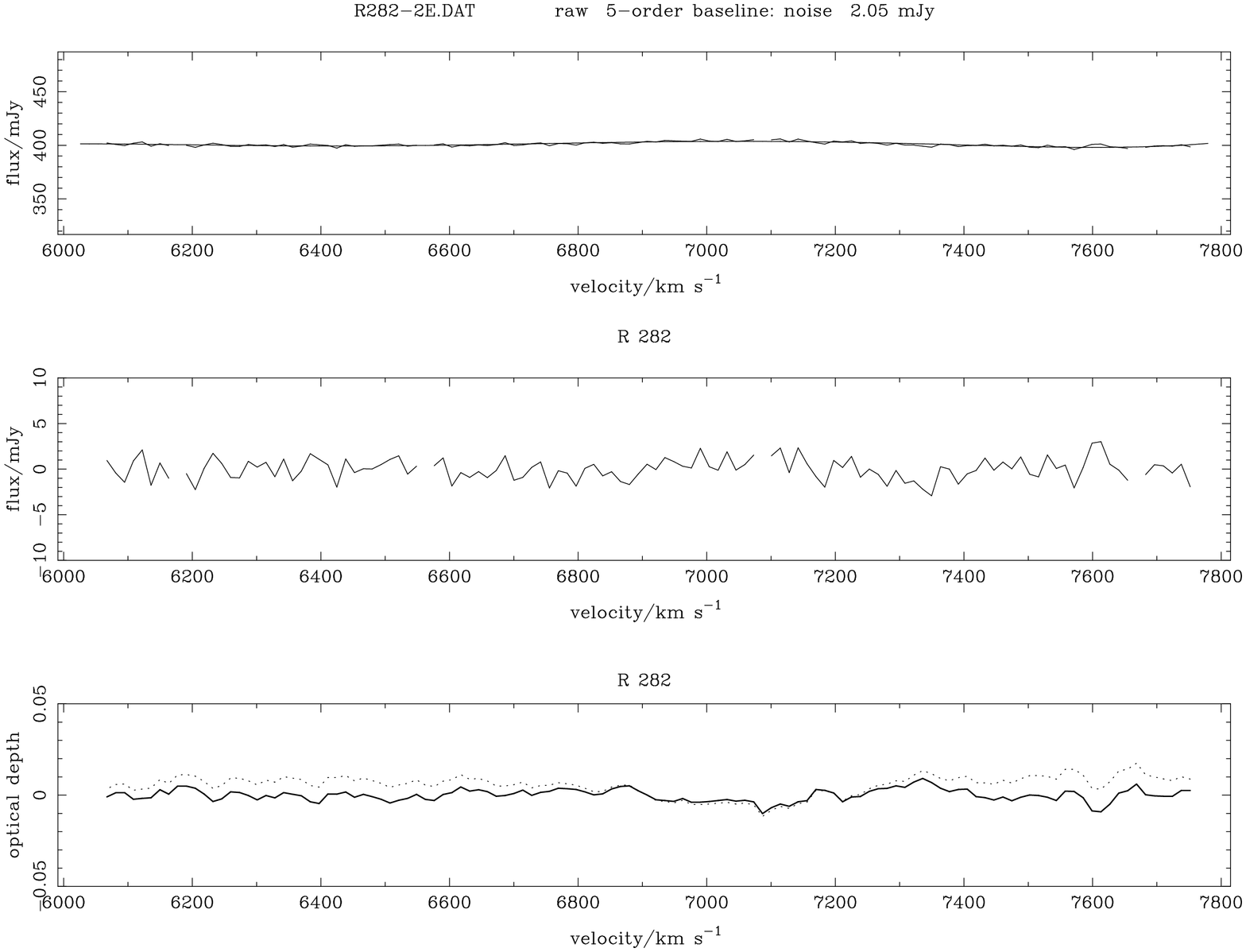}}
  \smallskip
\centerline{\includegraphics[angle=270,width=13.6cm,clip=]{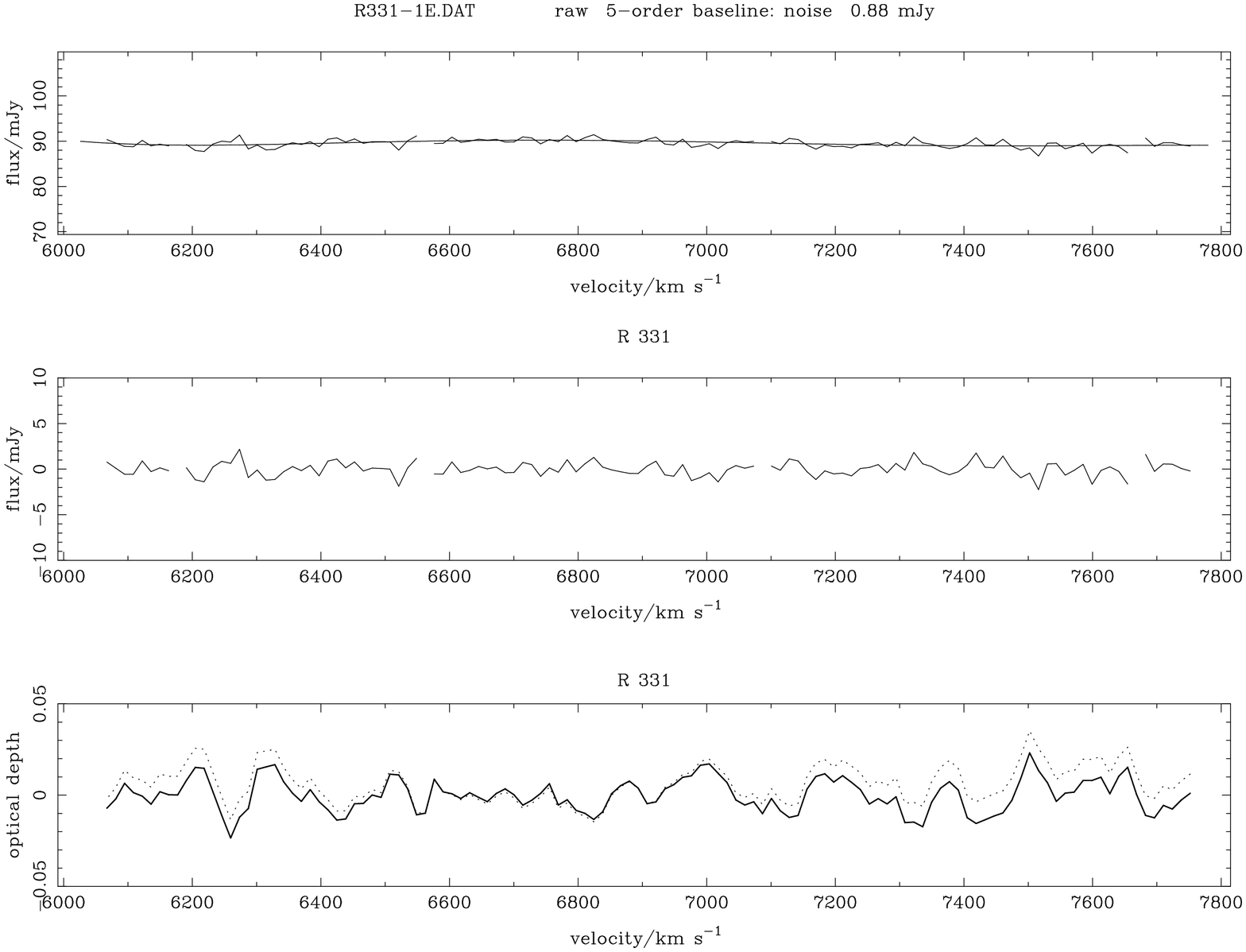}}
  \smallskip
\centerline{\includegraphics[angle=270,width=13.6cm,clip=]{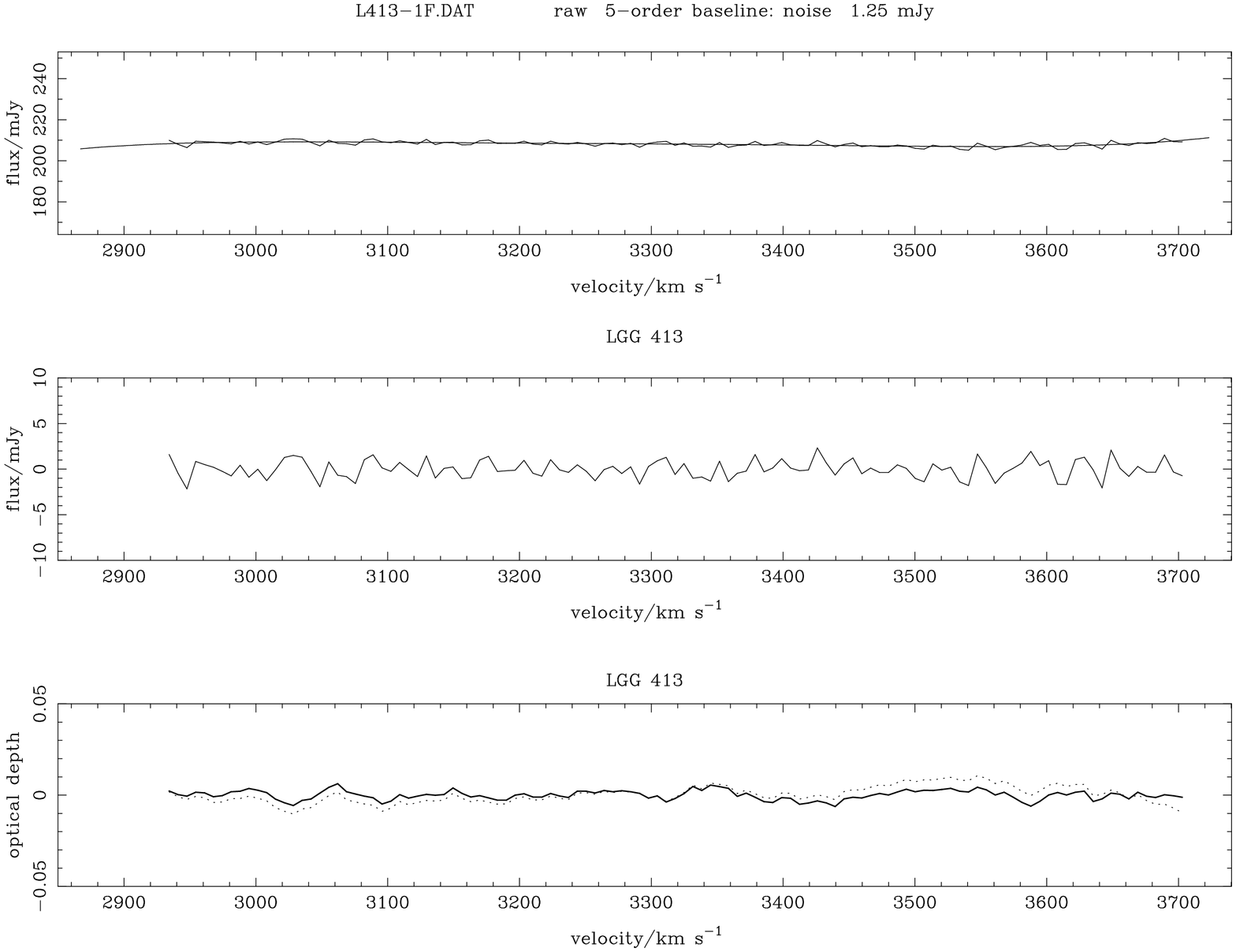}}
\caption{(continued).}
\end{figure}

\section{Results}\label{s:results}

The resulting {\Hi} spectra towards the observed, bright continuum sources are
shown in Fig.~\ref{f:results}. These results are shown in terms of optical
depth, $\tau$, where the observed flux density, $S_{\rm obs}$, is related to
the mean observed, continuum flux density, $S_{\rm cont}$ by
\begin{equation}
  S_{\rm obs} = S_{\rm cont} (1 - {\rm e}^{-\tau}),
\end{equation}
after smoothing the spectra to a resolution of 20 km s$^{-1}$ (cf.\ Manning
2002, who discusses the linewidths of clouds seen in Ly$\alpha$ absorption
studies, and Kalberla \& Huad 2006, who find typical dispersions of this value
for cold {\Hi} cores in Galactic High Velocity Clouds). These spectra do not
show any obvious absorption towards any of these sources. The spectra are
averages of both the left and right polarisations observed, and have had a
large scale (fifth order polynomial) baseline removed, to correct for residual
uncertainties in the bandpass calibrations. (The amplitude of this correction
was in all cases small, less than a few mJy.) For R 151, the results for two
nearby lines-of-sight, to different components of the NVSS source that was
resolved by these observations, are presented. A few channels at each end of
the observed band are excluded, as they are noisy due to large uncertainties in
the antenna-based bandpass corrections. Also, a few channels are omitted from
each of the 2003 January observations, due to correlator problems. These
results are also summarised in Table~\ref{t:results}, which gives the mean
observed flux density of the source, the rms deviation per channel -- after the
polynomial was removed -- and the signal-to-noise for each spectrum, together
with the rms of the smoothed spectra in terms of optical depth.

\begin{table*}
\centering
\begin{minipage}{13.0cm}
\caption{Results of {\Hi} absorption observations to background continuum
sources in galaxy groups. For R 151 results for two lines-of-sight are given,
as the NVSS source observed was resolved. The rms noise and signal-to-noise
values are for the observed channel widths (i.e.\ 6.7 and 13.2 km s$^{-1}$ for
observations observed with total bandwidths of 4 and 8~MHz respectively). The
optical depth rms is after smoothing to 20 km s$^{-1}$ resolution. $\theta$ is
the angular distance from the group centre.}\label{t:results}
\smallskip
\def\perbeam{{mJy~beam$^{-1}$}}
\tabcolsep3pt
\begin{tabular}{cccccccc}\hline
  Group    & \multicolumn{2}{c}{Source Position}             &          &    flux   &   rms      & signal &  optical \\
  ID       & \multicolumn{2}{c}{J2000}                       & $\theta$ &  density  &   noise    &   to   &  depth   \\
           &    h\quad m\quad s   & $\circ$\quad$'$\quad$''$ &  /arcmin & /\perbeam & /\perbeam  &  noise &   rms    \\ \hline
   R 97    &        10 48 38.5    &       $+$26 21 38        &     18   &    237    &   1.59     &   149  &  0.0044  \\
  R 110    &        11 00 29.7    &       $+$09 48 37        &     22   &    340    &   1.03     &   295  &  0.0036  \\  
  R 151    &        11 42 10.6    &       $+$10 06 17        &     22   &    236    &   1.38     &   171  &  0.0043  \\
           &        11 42 10.2    &       $+$10 06 13        &     22   &    187    &   1.26     &   148  &  0.0051  \\
  R 168    &        11 58 25.8    &       $+$24 50 19        &     18   &    468    &   1.45     &   323  &  0.0025  \\
  R 282    &        14 07 27.6    &       $+$09 17 40        &     24   &    402    &   1.17     &   344  &  0.0033  \\  
  R 331    &        15 07 30.4    &       $+$12 35 49        &     14   &     90    &   0.76     &   118  &  0.0088  \\  
 LGG 413   &        17 52 41.6    &       $+$24 36 03        &      7   &    208    &   0.97     &   215  &  0.0026  \\ \hline
\end{tabular}
\end{minipage}
\end{table*}

\section{Discussion and Conclusions}\label{s:conclusions}

Our observations constrain the mean column density along the sight-line. Using
\begin{equation}
  n_{\rm \subHi} \simeq 1.82 \times 10^{22} T_{\rm spin} \int \tau \, {\rm d}v
        \simeq 1.82 \times 10^{22} T_{\rm spin} \tau \, \Delta v
             \quad {\rm m}^{-2}
  \label{col-den}
\end{equation}
for velocities in km~s$^{-1}$, then the limits from our observations smoothed
to $\Delta v=20$ km~s$^{-1}$, with typically a $3\sigma$ limit of $\tau \la
0.01$, gives $n_{\rm \subHi} \la 3.6 \times 10^{23}$ m$^{-2}$, for a spin
temperature of 100~K. This limit is five times smaller that the peak {\Hi}
column densities detected in emission by Walter et al.\ (2005), for the
isolated {\Hi} cloud in the M81 group.

We can use this limit on the {\Hi} column density to constrain the mass
distribution of clouds in virial equilibrium in these galaxy groups. For a
projected area of the continuum source at the galaxy group $A$, and a
line-of-sight through the group $l$, then if the volume filling factor of {\Hi}
clouds in the galaxy group is $f$ then the volume of ``cloud'' we expect in our
surveyed volume is,
\begin{equation}
  V = A l f.
\end{equation}
If the number density of {\Hi} in our clouds is $\rho_{\rm \subHi}$ and our
column density limit is $n_{\rm \subHi}$ then;
\begin{equation}
  n_{\rm \subHi} = \frac{Alf \rho_{\rm \subHi}}{A} = lf\rho_{\rm \subHi}.
    \label{nHI}
\end{equation}
Now we can use the virial theorem to replace the dependence on $\rho_{\rm
\subHi}$ for a dependence on $r_{\rm c}$, the cloud radius. The virial
temperature is given by,
\begin{equation}
  T_{\rm vir} = 0.13 \frac{GM\mu m_{\rm p}}{k_{\rm B}r_{\rm c}}
   \label{Tvir}
\end{equation}
where $\mu$ is the mean particle mass in units of $m_{\rm p}$ (i.e.\ 1 for pure
{\Hi}). The cloud mass $M$ is simply
\begin{equation}
  M = \frac{4}{3}\pi r_{\rm c}^{3} \rho_{\rm \subHi} m_{\rm p}
\end{equation}
which gives
\begin{equation}
  T_{\rm vir} = 0.13 \frac{4G\pi m_{\rm p}^2 \mu}{3k_{\rm B}}
   \rho_{\rm \subHi} r_{\rm c}^2
\end{equation}
(for all parameters in S.I.\ units). Taking $T_{\rm vir} = T_{\rm spin}$, and
$\mu = 1$, then
\begin{equation}
  \rho_{\rm \subHi} r_{\rm c}^2 = 1.36 \times 10^{41} T_{\rm spin}
\end{equation}
Substituting in Eq.~\ref{nHI} for $\rho_{\rm \subHi}$ gives
\begin{equation}
  \frac{f}{r_{\rm c}^2} = \frac{n_{\rm \subHi}}{lT_{\rm spin}
    1.36 \times 10^{41}}
  \label{constraint}
\end{equation}
Combining the observational constraint of Eq.~\ref{col-den} with
Eq.~\ref{constraint} the dependence on spin temperature is eliminated, giving
\begin{equation}
  \frac{f}{r_{\rm c}^2}
      = \frac{1.82 \times 10^{22} \tau \Delta v}{1.36 \times 10^{41} l}
      = 1.34 \times 10^{-19} \frac{\tau \Delta v}{l}
\end{equation}
or, for $l$ and $r_{\rm c}$ are in units of pc,
\begin{equation}
  \frac{f}{r_{\rm c}^2}
     =0.0041 \frac{\tau \Delta v}{l}. \label{limit}
\end{equation}
Our ($3\sigma$) limits of $\tau$ of, typically, $0.01$ in 20~km~s$^{-1}$, for
$l=1$~Mpc, gives a statistical limit on $f/r_{h}^2$ of $8.3 \times 10^{-10}$.
For $f=1 \times 10^{-3}$ then $r_h=1$~kpc, or for $f=1 \times 10^{-6}$,
$r_h=35$~pc (although, of course, if the filling factor is low, a single line
of sight may not, by chance, cross a cloud). The sense of the limit is that
$f/r_{\rm c}^2$ is less than that given in Eq.\ref{limit} (i.e.\ for a given
filling factor we constrain the lower limit for the cloud radius). As we have
not considered a distribution of cloud sizes in this calculation $r_{\rm c}$ in
the above refers to the mean cloud radius.

Although the present observations provide useful constraints on any cool {\Hi}
in the galaxy groups studied, the results are limited due to the small number
of lines of sight probed. This is due to the low density on the sky of bright
enough background continuum sources. However, for more sensitive telescopes
(e.g.\ looking forward to the SKA, see Carilli \& Rawlings 2004), then this
technique of searching for hydrogen by absorption will gain in competitiveness
compared to searches for emission, as it will be possible to probe many more
lines of sight. Any increase in sensitivity will correspondingly directly
increase the detection sensitivity for emission constraints, whereas for
absorption studies the improvement is not only in terms of the direct
sensitivity to absorption towards any particular background source, but also
the number of accessible lines of sight will be increased. Consequently, the
overall usefulness of constraints provided by absorption studies will increase
more rapidly than for emission studies with more sensitive instruments (see
Kanekar \& Briggs 2004 for further discussion).

In conclusion, these GMRT observations provide additional constraints on the
amount of cool hydrogen clouds that could exist along 8 lines of sight through
7 galaxy groups. Although these results do not provide strong constraints on
any {\sl cool} {\Hi} in these galaxy groups, this method will, with the next
generation of telescopes such as the SKA, allow very many lines of sight to be
probed in a given group.

\section*{Acknowledgements}

We thank the staff of the GMRT who have made these observations possible. The
GMRT is run by the National Centre for Radio Astrophysics of the Tata Institute
of Fundamental Research, India.


\label{lastpage}

\end{document}